\documentclass[lettersize,journal]{IEEEtran}

\usepackage{amsmath,amsfonts,amssymb}
\usepackage{mathtools}

\usepackage{array}
\usepackage[caption=false,font=normalsize,labelfont=sf,textfont=sf]{subfig}
\usepackage{textcomp}
\usepackage{stfloats}
\usepackage{url}
\usepackage{verbatim}

\usepackage{graphicx}
\usepackage{float}

\usepackage{cite}

\usepackage{algorithm}
\usepackage{algpseudocode}

\usepackage{tabularray}
\UseTblrLibrary{booktabs}
\usepackage{bm}

\usepackage{physics}
\usepackage{siunitx}
\AtBeginDocument{\RenewCommandCopy\qty\SI}

\usepackage{placeins}
\usepackage{flafter}

\usepackage{hyperref}
\usepackage{orcidlink}

\newcommand*{\dif}{\mathop{}\!\mathrm{d}}

\begin{document}

\title{Medium-Induced Cross-Frequency Clutter Structure in Single-Snapshot FDA-MIMO-GPR With a Weak-Dispersion Criterion}

\author{Yisu Yan$^{\orcidlink{0009-0008-4022-3619}}$,~\IEEEmembership{Graduate Student Member,~IEEE,} Jifeng Guo$^{\orcidlink{0000-0002-9710-0045}}$ 

\thanks{Yisu Yan is with School of Astronautics, Harbin Institute of Technology, Heilongjiang, China
(e-mail: \href{mailto:23B918085@stu.hit.edu.cn}{23B918085@stu.hit.edu.cn}).}

\thanks{Jifeng Guo is with School of Astronautics, Harbin Institute of Technology, Heilongjiang, China (e-mail: \href{mailto:guojifeng@hit.edu.cn}{guojifeng@hit.edu.cn}).}

\thanks{This article is accompanied by supplementary material. The PDF document provides additional discussion on the well-posedness of the reference operator, the applicability range of the residual dielectric increment, the consistency of scene construction and Cole--Cole constitutive mapping, the closure of the main covariance term, the role of first-order covariance corrections, the truncation criterion for the main-term approximation, and supplementary ablation and zero-frequency-offset comparisons.}

\thanks{This work has been submitted to the IEEE for possible publication. Copyright may be transferred without notice, after which this version may no longer be accessible.}

}

\maketitle

\begin{abstract}
	This paper investigates the cross-frequency structure of background clutter induced by random dispersive media in single-snapshot FDA-MIMO-GPR. Representative media are modeled by the Cole--Cole formulation to relate dispersive constitutive behavior to the reference propagation environment and observation-domain statistics. A normalized incremental contrast function is introduced under a reference-medium framework, and a single-snapshot background-response expression with first-order propagation-kernel feedback is derived. Based on this expression, a cross-frequency coupling strength of the leading-order background covariance is defined. Numerical results show that, in weakly dispersive scenes, the proposed analysis remains consistent across constitutive mapping, the zeroth-order propagation skeleton, first-order distorted-Born truncation, propagation-kernel feedback, and single-channel response closure. The proposed metric distinguishes uncoupled and explicitly coupled constructions, remains stable under pure energy scaling, responds clearly to correlation length and relaxation-location parameters, and corresponds directly to the error of the frequency block-diagonal approximation. Additional experiments show that the resulting cross-frequency structure affects whitening and principal-subspace extraction. In scenes with pronounced relaxation, abrupt breakdown under strong perturbations and high-error plateaus indicate that the present theory is mainly applicable within the validity range of first-order feedback.
\end{abstract}


\section{Introduction}\label{sec:intro}

\IEEEPARstart{I}{n} many GPR applications, fine-scale non-target inhomogeneities in soil, weathered layers, or lunar regolith generate distributed background scattering. Such components may mask weak target echoes and alter the channel-domain statistical structure of the observations. Similar effects have been reported in GPR studies on landmine detection \cite{takahashi2010InfluenceSoilInhomogeneity,takahashi2012ModelingGPRClutter}, planetary shallow sounding \cite{feng2022ShallowRegolithStructure,zhang2025SubsoilStructureChangE6}, and shallow geological exploration \cite{salinasnaval2018GPRClutterAmplitude}. These observations indicate that background clutter induced by weak distributed medium fluctuations has both a clear physical origin and practical relevance.

This issue becomes more pronounced in single-snapshot FDA-MIMO-GPR. A single snapshot consists of one near-field joint space-frequency measurement, whose channel-domain structure is jointly shaped by array geometry, frequency-diverse channel allocation, interface transmission, and medium dispersion. The difficulty therefore arises from two sources: limited statistical support and explicit medium dependence in the clutter-generation mechanism itself. In this setting, background clutter involves not only frequencywise energy redistribution but also cross-frequency statistical coupling, which directly affects subsequent whitening, subspace extraction, and related structured processing.

Existing studies most closely related to the present work fall into three categories. The first concerns clutter statistical processing and detection for FDA/FDA-MIMO radar. This body of work has long examined clutter rank, degrees of freedom, and covariance structure under frequency-diverse waveforms and FDA-MIMO configurations. Existing studies have analyzed clutter-rank evolution, degree-of-freedom expansion, and the statistical dimensionality jointly determined by waveform and array parameters \cite{liu2016ClutterRanksFrequency,wang2022ClutterRankAnalysis}. On this basis, a variety of methods have been developed for clutter suppression and detection improvement, including multi-waveform suppression, tensorized STAP for space-time-range coupled observations, compressive sensing frameworks, and joint parameter optimization \cite{WEN2019280,wang2022RangeAmbiguousClutterSuppression,sun2024SpaceTimeRange,jia2026FDAMIMORadarParameter}. More recent studies continue along detector design, covariance control, and adaptive detection under prescribed statistical-background assumptions \cite{li2026AdaptiveTargetDetection}. In most of this literature, however, clutter is treated as a given interference field. Much less attention has been paid to how random dispersive subsurface fluctuations generate clutter structure in single-snapshot space-frequency observations.

The second category concerns forward modeling for GPR and MIMO-GPR, which is closer to subsurface propagation physics. Early FDA-MIMO-GPR models considered near-surface target detection, but usually assumed a homogeneous background, constant propagation parameters, and a scalar reflection coefficient for the target response \cite{liu2018DetectionSubsurfaceTarget}. More recent forward models for multiview, multistatic, and multifrequency GPR incorporate air--soil transmission, refracted propagation paths, and medium-dependent propagation kernels \cite{masoodi2024MultiviewMultistaticVs,gennarelli2026EffectEquivalentPermittivity}. Studies on planetary radar and shallow sounding have similarly shown that shallow observations are substantially affected by irregular weathered layers or lunar-regolith structure and cannot be interpreted solely through ideal layered interfaces \cite{feng2022ShallowRegolithStructure,zhang2025SubsoilStructureChangE6}. At the same time, many GPR clutter studies still focus mainly on post-echo suppression and normalization, such as signal-domain background suppression, spectral filtering, reflection-clutter removal, and clutter-distribution shaping \cite{kumar2025EnhancingSubsurfaceExploration,oliveira2021GPRClutterReflection,worthmann2021ClutterDistributionsTomographic}. Existing work has therefore not yet organized random medium dispersion, the reference propagation environment, and single-snapshot channel responses into a unified clutter-formation chain, nor has it clarified the resulting cross-frequency statistical structure in FDA-MIMO space-frequency observations.

The third category concerns constitutive dispersion and parameter uncertainty. Classical Debye and Cole--Cole models established that the complex permittivity of heterogeneous media is intrinsically frequency dependent. In particular, Cole and Cole provided the basis for distributed-relaxation descriptions, while Auty and Cole and later review studies further documented the dispersive behavior of representative media such as ice over different temperatures and frequency ranges \cite{cole1941DispersionAbsorptionDielectrics,auty1952DielectricPropertiesIce,holm2020TimeDomainCharacterization}. Such constitutive dispersion directly affects radar interpretation and inversion. Qin \emph{et al.} \cite{qin2023FullwaveformInversionGroundpenetrating}, for example, showed in GPR full-waveform inversion that neglecting attenuation and velocity dispersion significantly degrades the recovery of both conductive and scattering-dominated anomalies. Uncertainty-modeling studies have likewise emphasized that subsurface parameter fluctuations should be quantified explicitly rather than absorbed into deterministic inversion alone \cite{xie2021GPRUncertaintyModelling}. Similar conclusions also appear in material measurements and planetary scenarios: the electrical properties of lunar regolith and related samples vary with frequency, temperature, and water content \cite{strangway1972ElectricalPropertiesLunar,olhoeft1973ElectricalPropertiesLunar,olhoeft1974ElectricalPropertiesLunar}, while planetary simulants and porous media further vary with composition, porosity, and microstructure \cite{brouet2016CharacterizationPermittivityControlled,elshafie2013DielectricHardnessMeasurements}. Numerical simulation studies are also moving toward full-wave formulations with arbitrary complex permittivity \cite{majchrowska2021ModellingArbitraryComplex}. These studies, however, mainly remain at the level of dispersive-material characterization and its impact on inversion or simulation accuracy. They do not clarify how random constitutive dispersion is written through the propagation kernel into off-diagonal frequency covariance blocks and the resulting cross-frequency structure of single-snapshot FDA-MIMO-GPR observations.

Current studies therefore still lack a unified analytical route that starts from random dispersive constitutive behavior, passes through a reference propagation environment, and reaches cross-frequency structural quantities in the single-snapshot observation domain, while keeping the leading structure, higher-order corrections, and applicability boundaries distinct. To address this gap, this paper develops a layered analysis framework for random dispersive-media backgrounds in single-snapshot FDA-MIMO-GPR, from Cole--Cole constitutive mapping, the reference-medium propagation kernel, and the incremental background response of a single channel to cross-frequency quantities of the leading-order background covariance. The analysis focuses on background components induced by weak, distributed, non-target dispersive constitutive fluctuations. The expansion parameter is defined as the residual dielectric increment relative to the reference medium. This choice preserves the physical role of constitutive dispersion and propagation feedback while providing a controlled regime for structural analysis.

The main contributions of this paper are threefold. First, under a reference-medium framework, a normalized incremental contrast function is constructed and a single-snapshot background-response expression retaining first-order propagation-kernel feedback is derived, so that the zeroth-order propagation skeleton and the incremental feedback terms are written in a layered form. Second, a cross-frequency coupling strength of the leading-order background covariance is defined to measure the relative strength of off-diagonal frequency covariance blocks with respect to diagonal blocks, and this quantity is directly related to the error of the frequency block-diagonal approximation. Third, through consistency validation in weakly dispersive scenes and boundary diagnosis in strongly relaxing scenes, the paper identifies an analysis route whose leading structure is stable within its intended regime and whose applicability limits remain explicit.

This paper does not seek a high-accuracy closed-form covariance expression for all medium scenarios. Its goal is to identify the dominant cross-frequency structure written by random dispersive media into the single-snapshot observation domain and to construct structural quantities with diagnostic value. For weakly dispersive scenes, the numerical results provide strong consistency evidence. For strongly relaxing scenes, the emphasis shifts to boundary diagnosis and clarification of the applicability range. This positioning is consistent with both the theoretical assumptions and the numerical evidence.

The remainder of this paper is organized as follows. Section~\ref{sec:snapshot_model} first introduces the Cole--Cole medium model as the constitutive basis for subsequent modeling of random dispersive backgrounds. It then introduces the reference-medium propagation kernel and the distorted Born approximation to establish a background-response model for single-snapshot FDA-MIMO-GPR and its incremental representation relative to the reference state. Section~\ref{sec:cov} discusses the frequency-block structure of the leading-order background covariance, defines the cross-frequency coupling strength, and analyzes its relation to frequency-structure diagnosis. Section~\ref{sec:experiments} presents layered numerical validation, structural diagnosis, and minimal processing-consequence analysis to examine the applicability range and boundaries of the theoretical chain. Finally, Section~\ref{sec:conclusion} concludes the paper and discusses current limitations and possible extensions.

\section{Single-Snapshot Signal Model of FDA-MIMO-GPR}\label{sec:snapshot_model}

\subsection{Medium Characteristics Described by the Cole--Cole Model}

The Cole--Cole model \cite{cole1941DispersionAbsorptionDielectrics} is essentially a generalization of the Debye single-relaxation model. It uses a distribution parameter to extend a single relaxation time into a distribution around a central relaxation time. The classical expression of the complex permittivity in the Cole--Cole model can be written as

\begin{equation}
	\varepsilon_{r}^{*}(\omega)
	=
	\varepsilon_\infty
	+
	\frac{\varepsilon_s-\varepsilon_\infty}{1+(j\omega\tau)^{1-\alpha}}
	-\frac{j\sigma_{\mathrm{dc}}}{\omega\varepsilon_0}
	\label{eq:cole-cole}
\end{equation}

Eq.~\eqref{eq:cole-cole} contains three terms: the first two describe polarization relaxation, and the last term represents low-frequency loss caused by direct-current conduction. Accordingly, consider a general local medium at position $\bm{x}$. Let it be characterized by the local parameter vector

\begin{equation*}
	\bm{\mu}(\bm{x})
	=
	\left[
		\varepsilon_\infty(\bm{x}), \Delta\varepsilon(\bm{x}), \tau(\bm{x}), \alpha(\bm{x}), \sigma(\bm{x})
		\right]^\top
\end{equation*}
with elements representing the high-frequency limit, relaxation strength, characteristic relaxation time, spectral broadening, and conductive loss, respectively. The local complex permittivity is then written as

\begin{equation}
	\varepsilon_{c}(\omega,\bm{x})
	=
	\varepsilon_0
	\left[
		\varepsilon_\infty(\bm{x})
		+
		\frac{\Delta\varepsilon(\bm{x})}{1+(j\omega\tau(\bm{x}))^{1-\alpha(\bm{x})}}
		-
		\frac{j\sigma(\bm{x})}{\omega\varepsilon_0}
		\right]
	\label{eq:local_varepsilon}
\end{equation}

\subsection{Medium Decomposition and Perturbation}

The total medium is decomposed into a background medium and a local deviation relative to that background,

\begin{equation}
	\varepsilon_c(\omega,\bm{x})
	=
	\varepsilon_b(\omega,\bm{x})
	+
	\delta\varepsilon(\omega,\bm{x})
	\label{eq:bg+tg}
\end{equation}
where $\varepsilon_b(\omega,\bm{x})$ is the complex-permittivity field of the background medium, and $\delta\varepsilon(\omega,\bm{x})$ is the dielectric deviation of the local inhomogeneity relative to the background. The background medium $\varepsilon_b(\omega,\bm{x})$ satisfies \eqref{eq:cole-cole} and describes the nominal subsurface environment when strong deterministic structures are not explicitly extracted, thereby characterizing large-scale layering, slowly varying dispersion, and dissipative background effects.

A reference medium $\varepsilon_{\mathrm{ref}}$ is further introduced on top of $\varepsilon_b(\omega,\bm{x})$ as the working reference state, so as to absorb dominant deterministic strong interfaces, strongly reflecting layers, or known local strong scatterers. When no such dominant deterministic structures need to be absorbed separately, one may set $\varepsilon_{\mathrm{ref}}(\omega,\bm{x})=\varepsilon_b(\omega,\bm{x})$.

Let $\bm{\mu}_{b}(\bm{x})$ denote the nominal background-parameter vector of the scene, and let $\bm{\mu}_{\mathrm{ref}}(\bm{x})$ denote the reference-parameter vector after absorbing the dominant deterministic structures. Note that the local complex permittivity is essentially the mapping from Cole--Cole parameters to the constitutive response in the frequency domain,
\begin{equation*}
	\varepsilon_c(\omega,\bm{x})
	=
	\mathcal F_{\mathrm{CC}}\bigl(\omega;\bm{\mu}(\bm{x})\bigr)
\end{equation*}
The deviation of the actual local parameter vector relative to the reference state is then denoted by

\begin{equation}
	\delta\bm{\mu}(\bm{x})
	=
	\bm{\mu}(\bm{x})-\bm{\mu}_{\mathrm{ref}}(\bm{x})
	\label{eq:delta_mu_ref}
\end{equation}
Therefore, the dielectric increment relative to the reference state can be written as

\begin{equation}
	\Delta\varepsilon(\omega,\bm{x})
	=
	\mathcal F_{\mathrm{CC}}\bigl(\omega;\bm{\mu}_{\mathrm{ref}}(\bm{x})+\delta\bm{\mu}(\bm{x})\bigr)
	-
	\mathcal F_{\mathrm{CC}}\bigl(\omega;\bm{\mu}_{\mathrm{ref}}(\bm{x})\bigr)
	\label{eq:rel_dielectric_deviation}
\end{equation}

Specifically, $\varepsilon_b$ characterizes the physical baseline medium of the scene, $\varepsilon_{\mathrm{ref}}$ defines the working reference state for the subsequent propagation analysis, and $\Delta\varepsilon$ denotes the remaining dielectric increment relative to that reference state.

\subsection{Propagation Kernels in the Reference Medium}\label{subsec:propagation_kernels}

A nonmagnetic and locally isotropic medium is considered, namely $\mu(\bm{x})\approx \mu_0$, and the source current density is denoted by $\bm{J}_s$. The frequency-domain Maxwell system then yields the following second-order equation for the electric field $\bm{E}$:

\begin{equation}
	\nabla\times\nabla\times \bm{E}(\bm{x},\omega)
	-
	\omega^2\mu_0\varepsilon_c(\omega,\bm{x})\bm{E}(\bm{x},\omega)
	=
	-j\omega\mu_0 \bm{J}_s(\bm{x},\omega)
	\label{eq:bmE_2order}
\end{equation}
Accordingly, the local Maxwell operator controlled by the medium parameters is introduced as

\begin{equation}
	\mathcal{L}(\omega;\varepsilon_c)\bm{E}
	:=
	\nabla\times\nabla\times \bm{E}
	-
	\omega^2\mu_0\varepsilon_c(\omega,\bm{x})\bm{E}
	\label{eq:cal_maxwell}
\end{equation}

Through operator \eqref{eq:cal_maxwell}, the medium parameters directly determine the basic operator governing field propagation and scattering. Under the reference medium $\varepsilon_{\mathrm{ref}}(\omega,\bm x)$, the corresponding reference Maxwell operator is denoted by
\begin{equation}
	\mathcal L_{\mathrm{ref}}(\omega)\bm E
	:=
	\nabla\times\nabla\times\bm E
	-
	\omega^2\mu_0\varepsilon_{\mathrm{ref}}(\omega,\bm x)\bm E
	\label{eq:cal_ref_maxwell}
\end{equation}
For each frequency in the operating band, the reference scattering problem is assumed to admit a unique radiating solution under the corresponding outgoing radiation condition. The reference Green tensor $\bm G_{\mathrm{ref}}$ is therefore well defined and satisfies

\begin{equation}
	\mathcal L_{\mathrm{ref}}(\omega)\bm G_{\mathrm{ref}}(\bm x,\bm x';\omega)
	=
	\bm I\,\delta(\bm x-\bm x')
	\label{eq:green_ref}
\end{equation}
where $\bm I$ is the second-order identity tensor and $\delta(\cdot)$ is the Dirac delta function. If the source current density of the $n$th transmit channel at frequency $\omega_n$ is $\bm J_{s,n}(\bm x,\omega_n)$, then the incident field in the reference medium can be written as

\begin{equation}
	\bm E^{\mathrm{inc}}_{n,\mathrm{ref}}(\bm x,\omega_n)
	=
	-j\omega_n\mu_0
	\int
	\bm G_{\mathrm{ref}}(\bm x,\bm x';\omega_n)\,
	\bm J_{s,n}(\bm x',\omega_n)
	\dif\bm x'
	\label{eq:Einc_green_ref}
\end{equation}

Let $\bm p_{t,n}(\bm x,\omega_n)$ denote the equivalent transmitter-side polarization projection near the scattering point, $\bm w_m(\bm x,\omega_n)$ the corresponding receive-side test function, and $\bm p_r(\bm x,\omega_n)$ the local equivalent polarization direction of the scattering source. The transmit and receive propagation kernels under the reference state are then defined, respectively, as

\begin{equation}
	G_t^{(0)}(\bm x,\bm s_n;\omega_n)
	:=
	\bm p_{t,n}^{\mathrm H}(\bm x,\omega_n)\,
	\bm E^{\mathrm{inc}}_{n,\mathrm{ref}}(\bm x,\omega_n)
	\label{eq:Gt_ref_def}
\end{equation}
and

\begin{equation}
	G_r^{(0)}(\bm r_m,\bm x;\omega_n)
	:=
	\int
	\bm w_m^{\mathrm H}(\bm x',\omega_n)\,
	\bm G_{\mathrm{ref}}(\bm x',\bm x;\omega_n)\,
	\bm p_r(\bm x,\omega_n)
	\dif\bm x'
	\label{eq:Gr_ref_def}
\end{equation}
$G_t^{(0)}$ and $G_r^{(0)}$ are essentially scalarized forms of the reference Green tensor after projection through source distribution, polarization mode, aperture weighting, and receive test functions.

\subsection{Distorted Born Linearization of the Increment}\label{subsec:dba_linearization}

Once the background baseline $\varepsilon_b$ has been specified and the reference medium $\varepsilon_{\mathrm{ref}}$ has also been fixed at the constitutive level, the corresponding reference field $\bm E^{\mathrm{ref}}$ satisfies

\begin{equation}
	\mathcal L_{\mathrm{ref}}(\omega)\bm E^{\mathrm{ref}}(\bm x,\omega)
	=
	-j\omega\mu_0\bm J_s(\bm x,\omega)
	\label{eq:field_ref_operator}
\end{equation}

Let the residual dielectric increment of the total medium relative to the reference medium be denoted by
\begin{equation}
	\Delta\varepsilon(\omega,\bm x)
	:=
	\varepsilon_c(\omega,\bm x)-\varepsilon_{\mathrm{ref}}(\omega,\bm x)
	\label{eq:delta_eps_increment}
\end{equation}
which can be further written as \eqref{eq:rel_dielectric_deviation} through the Cole--Cole parameter mapping. The total field can then be written around the reference medium as

\begin{multline}
	\bm E(\bm x,\omega)
	=
	\bm E^{\mathrm{ref}}(\bm x,\omega) \\
	+
	\omega^2\mu_0
	\int_D
	\bm G_{\mathrm{ref}}(\bm x,\bm x';\omega)\,
	\Delta\varepsilon(\omega,\bm x')\,
	\bm E(\bm x',\omega)
	\dif\bm x'
	\label{eq:field_dba_exact}
\end{multline}

In \eqref{eq:field_dba_exact}, the dominant deterministic propagation distortion, strong-interface effects, and known strong scattering structures have already been absorbed into the reference medium. For the local residual increment relative to the reference state, the \emph{distorted Born approximation} is adopted by replacing the total field in the integral term with the reference field, namely,

\begin{multline}
	\bm E(\bm x,\omega)
	\approx
	\bm E^{\mathrm{ref}}(\bm x,\omega) \\
	+
	\omega^2\mu_0
	\int_D
	\bm G_{\mathrm{ref}}(\bm x,\bm x';\omega)\,
	\Delta\varepsilon(\omega,\bm x')\,
	\bm E^{\mathrm{ref}}(\bm x',\omega)
	\dif\bm x'
	\label{eq:field_dba_approx}
\end{multline}

Under the same reference state, the first-order response of the propagation kernels to the residual dielectric increment is given by the perturbation of the reference Green tensor. The true Green tensor admits the following first-order expansion around the reference state:

\begin{gather}
	\bm G(\bm x,\bm x';\omega)
	=
	\bm G_{\mathrm{ref}}(\bm x,\bm x';\omega)
	+
	\delta\bm G(\bm x,\bm x';\omega)
	+
	O\!\left(\|\Delta\varepsilon\|^2\right)
	\label{eq:green_perturb} \\
	\delta\bm G(\bm x,\bm x';\omega)
	=
	\omega^2\mu_0
	\int_D
	\bm G_{\mathrm{ref}}(\bm x,\bm z;\omega)\,
	\Delta\varepsilon(\omega,\bm z)\,
	\bm G_{\mathrm{ref}}(\bm z,\bm x';\omega)
	\dif\bm z
	\label{eq:delta_green}
\end{gather}
Accordingly, the incident field can also be written as

\begin{gather}
	\bm E^{\mathrm{inc}}_{n}(\bm x,\omega_n)
	=
	\bm E^{\mathrm{inc}}_{n,\mathrm{ref}}(\bm x,\omega_n)
	+
	\delta\bm E^{\mathrm{inc}}_{n}(\bm x,\omega_n)
	+
	O\!\left(\|\Delta\varepsilon\|^2\right)
	\label{eq:Einc_expand} \\
	\delta\bm E^{\mathrm{inc}}_{n}(\bm x,\omega_n)
	=
	\omega_n^2\mu_0
	\int_D
	\bm G_{\mathrm{ref}}(\bm x,\bm z;\omega_n)\,
	\Delta\varepsilon(\omega_n,\bm z)\,
	\bm E^{\mathrm{inc}}_{n,\mathrm{ref}}(\bm z,\omega_n)
	\dif\bm z
	\label{eq:delta_Einc}
\end{gather}
The transmit and receive propagation kernels then admit the expansions

\begin{gather}
	G_t(\bm x,\bm s_n;\omega_n)
	=
	G_t^{(0)}(\bm x,\bm s_n;\omega_n)
	+
	\delta G_t(\bm x,\bm s_n;\omega_n)
	+
	O\!\left(\|\Delta\varepsilon\|^2\right)
	\label{eq:Gt_expand} \\
	G_r(\bm r_m,\bm x;\omega_n)
	=
	G_r^{(0)}(\bm r_m,\bm x;\omega_n)
	+
	\delta G_r(\bm r_m,\bm x;\omega_n)
	+
	O\!\left(\|\Delta\varepsilon\|^2\right)
	\label{eq:Gr_expand} \\
	\delta G_t(\bm x,\bm s_n;\omega_n)
	:=
	\bm p_{t,n}^{\mathrm H}(\bm x,\omega_n)\,
	\delta\bm E^{\mathrm{inc}}_{n}(\bm x,\omega_n)
	\label{eq:delta_Gt_def} \\
	\delta G_r(\bm r_m,\bm x;\omega_n)
	:=
	\int
	\bm w_m^{\mathrm H}(\bm z,\omega_n)\,
	\delta\bm G(\bm z,\bm x;\omega_n)\,
	\bm p_r(\bm x,\omega_n)
	\dif\bm z
	\label{eq:delta_Gr_def}
\end{gather}
On this basis, a normalized incremental contrast function $\xi$ relative to the reference medium is defined by

\begin{equation}
	\begin{aligned}
		\xi(\bm{x},\omega)
		 & :=
		\frac{\Delta\varepsilon(\omega,\bm{x})}{\varepsilon_{\mathrm{ref}}(\omega,\bm{x})} \\
		 & =
		\frac{
		\mathcal F_{\mathrm{CC}}\bigl(\omega;\bm{\mu}_{\mathrm{ref}}(\bm{x})+\delta\bm{\mu}(\bm{x})\bigr)
		-
		\mathcal F_{\mathrm{CC}}\bigl(\omega;\bm{\mu}_{\mathrm{ref}}(\bm{x})\bigr)
		}{
		\varepsilon_{\mathrm{ref}}(\omega,\bm{x})
		}
	\end{aligned}
	\label{eq:xi}
\end{equation}
to represent the incremental scattering strength above the reference state. It is assumed that $\varepsilon_{\mathrm{ref}}(\omega,\bm x)$ is bounded and nonzero over the operating band and region, so that the normalization in \eqref{eq:xi} is well defined.

In channelized form, the echo of the $n$th frequency-diverse transmit channel and the $m$th receive channel at the selected angular frequency $\omega_n$ is written as

\begin{equation}
	\begin{aligned}
		y_{mn}(\omega_n)
		= &
		\int_D
		G_r(\bm r_m,\bm x;\omega_n)\,
		\xi(\bm x,\omega_n)\,
		G_t(\bm x,\bm s_n;\omega_n)
		\dif\bm x \\
		  & +
		n_{mn}(\omega_n)
	\end{aligned}
	\label{eq:echo_n_m_omegan}
\end{equation}

Substituting \eqref{eq:Gt_expand}--\eqref{eq:Gr_expand} into \eqref{eq:echo_n_m_omegan} and retaining the first-order perturbation terms of the propagation kernels yields

\begin{equation}
	\begin{aligned}
		y_{mn}(\omega_n)
		\approx &
		\int_D
		G_r^{(0)}(\bm r_m,\bm x;\omega_n)\,
		\xi(\bm x,\omega_n)\,
		G_t^{(0)}(\bm x,\bm s_n;\omega_n)
		\dif\bm x   \\
		        & +
		\int_D
		\delta G_r(\bm r_m,\bm x;\omega_n)\,
		\xi(\bm x,\omega_n)\,
		G_t^{(0)}(\bm x,\bm s_n;\omega_n)
		\dif\bm x   \\
		        & +
		\int_D
		G_r^{(0)}(\bm r_m,\bm x;\omega_n)\,
		\xi(\bm x,\omega_n)\,
		\delta G_t(\bm x,\bm s_n;\omega_n)
		\dif\bm x   \\
		        & +
		n_{mn}(\omega_n)
	\end{aligned}
	\label{eq:echo_n_m_omegan_1st_kernel}
\end{equation}
The first term on the right-hand side represents the leading incremental background response under the reference propagation environment. The second and third terms account for the first-order feedback of the residual medium increment on the receive-side and transmit-side propagation kernels, respectively, thereby introducing illumination-path and echo-path redistribution. Equation~\eqref{eq:echo_n_m_omegan_1st_kernel} therefore yields a semi-nonlinear second-order correction model, in which $\xi$ is retained exactly, whereas $\delta G_r$ and $\delta G_t$ are kept only to first order.

\section{Covariance Structure of Medium-Induced Responses}\label{sec:cov}

\subsection{Covariance Expression of the Medium-Induced Background Response}\label{subsec:clutter_covariance}

Removing the noise term from \eqref{eq:echo_n_m_omegan_1st_kernel}, the corresponding medium-induced background response is denoted by
\begin{equation}
	\begin{aligned}
		c_{mn}(\omega_n)
		 & :=
		y_{mn}(\omega_n)-n_{mn}(\omega_n) \\
		 & \approx
		c_{mn}^{(0)}(\omega_n)
		+
		c_{mn}^{(1)}(\omega_n)
	\end{aligned}
	\label{eq:clutter_component_decomp}
\end{equation}
where the leading incremental background response in the reference propagation environment is
\begin{equation}
	c_{mn}^{(0)}(\omega_n)
	:=
	\int_D
	G_r^{(0)}(\bm r_m,\bm x;\omega_n)\,
	\xi(\bm x,\omega_n)\,
	G_t^{(0)}(\bm x,\bm s_n;\omega_n)
	\dif\bm x
	\label{eq:clutter_main_term}
\end{equation}
and the correction induced by first-order feedback of the propagation kernels is
\begin{gather}
	c_{mn}^{(1)}(\omega_n)
	:=
	c_{mn}^{(1,r)}(\omega_n)
	+
	c_{mn}^{(1,t)}(\omega_n)
	\label{eq:clutter_feedback_decomp}
	\\
	c_{mn}^{(1,r)}(\omega_n)
	:=
	\int_D
	\delta G_r(\bm r_m,\bm x;\omega_n)\,
	\xi(\bm x,\omega_n)\,
	G_t^{(0)}(\bm x,\bm s_n;\omega_n)
	\dif\bm x
	\label{eq:clutter_feedback_r}
	\\
	c_{mn}^{(1,t)}(\omega_n)
	:=
	\int_D
	G_r^{(0)}(\bm r_m,\bm x;\omega_n)\,
	\xi(\bm x,\omega_n)\,
	\delta G_t(\bm x,\bm s_n;\omega_n)
	\dif\bm x.
	\label{eq:clutter_feedback_t}
\end{gather}
To write the first-order feedback term explicitly as a bilinear functional of $\xi$, note that
\begin{equation}
	\Delta\varepsilon(\omega,\bm x)
	=
	\varepsilon_{\mathrm{ref}}(\omega,\bm x)\,\xi(\bm x,\omega)
	\label{eq:delta_eps_xi_relation}
\end{equation}
Substituting \eqref{eq:delta_eps_xi_relation} into \eqref{eq:delta_Einc} yields
\begin{equation}
	\delta G_t(\bm x,\bm s_n;\omega_n)
	=
	\int_D
	h_n^{(t)}(\bm x,\bm z;\omega_n)\,
	\xi(\bm z,\omega_n)
	\dif\bm z
	\label{eq:delta_Gt_kernel_xi}
\end{equation}
where
\begin{multline}
	h_n^{(t)}(\bm x,\bm z;\omega_n)
	:= \\
	\omega_n^2\mu_0\,
	\bm p_{t,n}^{\mathrm H}(\bm x,\omega_n)\,
	\bm G_{\mathrm{ref}}(\bm x,\bm z;\omega_n)\,
	\varepsilon_{\mathrm{ref}}(\omega_n,\bm z)\,
	\bm E^{\mathrm{inc}}_{n,\mathrm{ref}}(\bm z,\omega_n)
	\label{eq:ht_def}
\end{multline}
Similarly, substituting \eqref{eq:delta_eps_xi_relation} into \eqref{eq:delta_green} and \eqref{eq:delta_Gr_def} yields
\begin{equation}
	\delta G_r(\bm r_m,\bm x;\omega_n)
	=
	\int_D
	h_{mn}^{(r)}(\bm x,\bm z;\omega_n)\,
	\xi(\bm z,\omega_n)
	\dif\bm z
	\label{eq:delta_Gr_kernel_xi}
\end{equation}
where
\begin{multline}
	h_{mn}^{(r)}(\bm x,\bm z;\omega_n)
	:= \\
	\omega_n^2\mu_0
	\int
	\bm w_m^{\mathrm H}(\bm z',\omega_n)\,
	\bm G_{\mathrm{ref}}(\bm z',\bm z;\omega_n)\,
	\varepsilon_{\mathrm{ref}}(\omega_n,\bm z) \\
	\bm G_{\mathrm{ref}}(\bm z,\bm x;\omega_n)\,
	\bm p_r(\bm x,\omega_n)
	\dif\bm z'
	\label{eq:hr_def}
\end{multline}

Accordingly, define the second-order feedback kernel
\begin{multline}
	b_{mn}(\bm x,\bm z;\omega_n)
	:= \\
	h_{mn}^{(r)}(\bm x,\bm z;\omega_n)\,
	G_t^{(0)}(\bm x,\bm s_n;\omega_n) \\
	+
	G_r^{(0)}(\bm r_m,\bm x;\omega_n)\,
	h_n^{(t)}(\bm x,\bm z;\omega_n)
	\label{eq:bmn_def}
\end{multline}
Then \eqref{eq:clutter_feedback_decomp} can be combined into
\begin{equation}
	c_{mn}^{(1)}(\omega_n)
	=
	\int_D\int_D
	b_{mn}(\bm x,\bm z;\omega_n)\,
	\xi(\bm x,\omega_n)\,
	\xi(\bm z,\omega_n)
	\dif\bm z\,\dif\bm x
	\label{eq:clutter_feedback_bilinear}
\end{equation}

Therefore, under the present semi-nonlinear model, the single-channel background response consists of a linear leading term in $\xi$ and a quadratic feedback term. The latter arises from the product of the first-order feedback of the propagation kernels to the residual medium increment and the original incremental scattering term.

Stack the background responses of all channels in a fixed order as
\begin{equation}
	\bm c
	:=
	\big[
	c_{11}(\omega_1),\dots,c_{M1}(\omega_1),
	c_{12}(\omega_2),\dots,c_{MN}(\omega_N)
	\big]^{\mathrm T}
	\in\mathbb C^{MN}
	\label{eq:clutter_vector_def}
\end{equation}
Then the snapshot background-response covariance is defined by
\begin{equation}
	\bm R_c
	:=
	\operatorname{Cov}(\bm c)
	=
	\mathbb E\!\left[
		\bigl(\bm c-\mathbb E[\bm c]\bigr)
		\bigl(\bm c-\mathbb E[\bm c]\bigr)^{\mathrm H}
		\right]
	\label{eq:Rc_def}
\end{equation}

If the receiver noise is further assumed to be independent of the medium randomness, then the snapshot observation covariance satisfies
\begin{gather}
	\bm R_y
	=
	\bm R_c+\bm R_n
	\label{eq:Ry_Rc_Rn}
	\\
	\bm R_n:=\operatorname{Cov}(\bm n)
\end{gather}

When the relevant central moments exist and expectation and integration can be interchanged, the following decomposition of $\bm R_c$ is obtained. Let
\begin{equation}
	\xi_n(\bm x):=\xi(\bm x,\omega_n),
	\quad
	\tilde\xi_n(\bm x):=\xi_n(\bm x)-\mathbb E[\xi_n(\bm x)]
	\label{eq:xi_centered}
\end{equation}
Accordingly, $\bm c$ can be written as
\begin{equation}
	\bm c
	\approx
	\bm c^{(0)}+\bm c^{(1)}
	\label{eq:c_vector_decomp}
\end{equation}
where $\bm c^{(0)}$ and $\bm c^{(1)}$ are formed by stacking the channelwise responses in \eqref{eq:clutter_main_term} and \eqref{eq:clutter_feedback_bilinear}, respectively. Then one has
\begin{gather}
	\bm R_c
	\approx
	\bm R_c^{(0)}
	+
	\bm R_c^{(0,1)}
	+
	\bigl(\bm R_c^{(0,1)}\bigr)^{\mathrm H}
	+
	\bm R_c^{(1)}
	\label{eq:Rc_decomp}
	\\
	\bm R_c^{(0)}
	:=
	\operatorname{Cov}\!\bigl(\bm c^{(0)}\bigr)
	\\
	\bm R_c^{(0,1)}
	:=
	\operatorname{Cov}\!\bigl(\bm c^{(0)},\bm c^{(1)}\bigr)
	\\
	\bm R_c^{(1)}
	:=
	\operatorname{Cov}\!\bigl(\bm c^{(1)}\bigr)
	\label{eq:Rc_terms}
\end{gather}

For any two channel indices $p=(m,n)$ and $q=(m',n')$, define the second central-moment kernel of the contrast field as
\begin{equation}
	C_{\xi}^{(2)}
	\bigl[
		(\bm x,\omega_n),(\bm x',\omega_{n'})
		\bigr]
	:=
	\mathbb E\!\left[
		\tilde\xi_n(\bm x)\,
		\tilde\xi_{n'}(\bm x')^{*}
		\right]
	\label{eq:Cxi2_def}
\end{equation}
Then the matrix entry of the leading-order covariance is
\begin{multline}
	\bigl[\bm R_c^{(0)}\bigr]_{p,q}
	=
	\int_D\int_D
	G_r^{(0)}(\bm r_m,\bm x;\omega_n)\,
	C_{\xi}^{(2)}
	\bigl[
		(\bm x,\omega_n),(\bm x',\omega_{n'})
		\bigr]                \\
	\times
	\overline{G_r^{(0)}(\bm r_{m'},\bm x';\omega_{n'})}\,
	G_t^{(0)}(\bm x,\bm s_n;\omega_n)\,
	\overline{G_t^{(0)}(\bm x',\bm s_{n'};\omega_{n'})}
	\dif\bm x\,\dif\bm x'
	\label{eq:Rc0_entry}
\end{multline}

Equation~\eqref{eq:Rc0_entry} shows that, before further treatment of propagation-kernel feedback, the leading background covariance is determined entirely by the reference propagation kernels and the second central-moment kernel of the contrast field. The mixed covariance $\bm R_c^{(0,1)}$ is governed by the third central moment of the contrast field, whereas the pure-feedback covariance $\bm R_c^{(1)}$ is governed by the fourth. Their explicit forms follow from termwise expansion of \eqref{eq:clutter_feedback_bilinear} and are omitted for brevity.

Under the present semi-nonlinear model, the leading covariance is therefore propagated from second-order statistics, while propagation-kernel feedback contributes higher-order mixed and pure-feedback corrections. The subsequent analysis focuses on $\bm R_c^{(0)}$, whose explicit propagation form is given by \eqref{eq:Rc0_entry}, since this term most directly captures the dominant second-order structure written by the random dispersive medium into the single-snapshot observation covariance.

\subsection{Cross-Frequency Coupling Strength of the Leading-Order Clutter Covariance}\label{subsec:cross_frequency_coupling}

Consider the leading-order background covariance $\bm R_c^{(0)}$. According to frequency, rewrite the leading-order background response in \eqref{eq:clutter_vector_def} in block form as
\begin{equation}
	\bm c_n^{(0)}
	:=
	\bigl[
	c_{1n}^{(0)}(\omega_n),\dots,c_{Mn}^{(0)}(\omega_n)
	\bigr]^{\mathrm T}
	\in\mathbb C^M
	\label{eq:c0_block_def}
\end{equation}
Then one has
\begin{equation}
	\bm c^{(0)}
	=
	\bigl[
	(\bm c_1^{(0)})^{\mathrm T},\dots,(\bm c_N^{(0)})^{\mathrm T}
	\bigr]^{\mathrm T}
	\label{eq:c0_block_stack}
\end{equation}

Further define the frequency blocks of the leading-order background covariance as
\begin{equation}
	\bm R_{c,nn'}^{(0)}
	:=
	\operatorname{Cov}\!\bigl(\bm c_n^{(0)},\bm c_{n'}^{(0)}\bigr)
	\in\mathbb C^{M\times M}
	\label{eq:Rc0_block_def}
\end{equation}
Then $\bm R_c^{(0)}$ can be written as
\begin{equation}
	\bm R_c^{(0)}
	=
	\begin{bmatrix}
		\bm R_{c,11}^{(0)} & \cdots & \bm R_{c,1N}^{(0)} \\
		\vdots             & \ddots & \vdots             \\
		\bm R_{c,N1}^{(0)} & \cdots & \bm R_{c,NN}^{(0)}
	\end{bmatrix}
	\label{eq:Rc0_block_matrix}
\end{equation}

To express each frequency block compactly, define the frequency-coupling kernel matrix
\begin{equation}
	\bm \Gamma_{\xi}(\bm x,\bm x')
	:=
	\Bigl[
		C_{\xi}^{(2)}
		\bigl[
			(\bm x,\omega_n),(\bm x',\omega_{n'})
			\bigr]
		\Bigr]_{n,n'=1}^{N}
	\label{eq:Gamma_xi_def}
\end{equation}
and the channel-response vector corresponding to the $n$th frequency point as
\begin{equation}
	[\bm a_n(\bm x)]_m
	:=
	G_r^{(0)}(\bm r_m,\bm x;\omega_n)\,
	G_t^{(0)}(\bm x,\bm s_n;\omega_n)
	\label{eq:a_n_def}
\end{equation}
Then, from \eqref{eq:clutter_main_term} and \eqref{eq:Cxi2_def}, one obtains
\begin{equation}
	\bm c_n^{(0)}
	=
	\int_D
	\bm a_n(\bm x)\,
	\xi_n(\bm x)
	\dif\bm x
	\label{eq:c0_block_compact}
\end{equation}
and
\begin{multline}
	\bm R_{c,nn'}^{(0)}
	= \\
	\int_D\int_D
	\bm a_n(\bm x)\,
	C_{\xi}^{(2)}
	\bigl[
		(\bm x,\omega_n),(\bm x',\omega_{n'})
		\bigr]\,
	\bm a_{n'}(\bm x')^{\mathrm H}
	\dif\bm x\,\dif\bm x'
	\label{eq:Rc0_block_entry}
\end{multline}

Equation~\eqref{eq:Rc0_block_entry} shows that the off-diagonal frequency blocks of $\bm R_c^{(0)}$ are obtained by projecting the off-diagonal entries of $\bm \Gamma_{\xi}(\bm x,\bm x')$ through the reference propagation kernels. They therefore characterize the cross-frequency statistical coupling induced by medium randomness in the observation domain. Accordingly, define the cross-frequency coupling strength of the leading-order background covariance as
\begin{equation}
	\chi_f\bigl(\bm R_c^{(0)}\bigr)
	:=
	\frac{
		\sum_{n\neq n'}
		\bigl\|
		\bm R_{c,nn'}^{(0)}
		\bigr\|_F^2
	}{
		\sum_{n=1}^{N}
		\bigl\|
		\bm R_{c,nn}^{(0)}
		\bigr\|_F^2
	}
	\label{eq:chi_f_def}
\end{equation}
where $\|\cdot\|_F$ denotes the Frobenius norm. If the denominator is nonzero, then
\begin{equation}
	\chi_f\bigl(\bm R_c^{(0)}\bigr)=0
	\iff
	\bm R_{c,nn'}^{(0)}=\bm 0,
	\quad
	\forall\, n\neq n'
	\label{eq:chi_f_zero}
\end{equation}

Further, if for any $n\neq n'$ it holds that
\begin{equation}
	C_{\xi}^{(2)}
	\bigl[
		(\bm x,\omega_n),(\bm x',\omega_{n'})
		\bigr]
	=
	0,
	\quad
	\forall\, \bm x,\bm x'\in D
	\label{eq:Cxi2_offdiag_zero}
\end{equation}
then, from \eqref{eq:Rc0_block_entry}, one obtains
\begin{equation}
	\bm R_{c,nn'}^{(0)}
	=
	\bm 0,
	\quad
	\forall\, n\neq n'
	\label{eq:Rc0_block_zero}
\end{equation}
and hence
\begin{equation}
	\chi_f\bigl(\bm R_c^{(0)}\bigr)=0
	\label{eq:chi_f_zero_sufficient}
\end{equation}

Equations~\eqref{eq:Cxi2_offdiag_zero}--\eqref{eq:chi_f_zero_sufficient} show that $\chi_f\bigl(\bm R_c^{(0)}\bigr)$ quantifies the relative strength of off-diagonal frequency blocks in the leading-order background covariance. It therefore indicates whether cross-frequency second-order coupling in the contrast field has been written into the leading-order observation covariance through the reference propagation kernels. Large values imply substantial retention of cross-frequency coupling in the snapshot background response, whereas values close to zero justify a frequency block-diagonal approximation.

\subsection{Implications of Cross-Frequency Coupling Analysis}\label{subsec:meaning_of_cross_frequency_coupling}

The cross-frequency coupling strength of the leading-order background covariance distinguishes between two covariance-formation mechanisms. One is dominated by frequencywise redistribution of marginal intensity, so that the observation covariance is structured mainly through frequencywise coloring. The other includes genuine second-order coupling across frequency channels, in which case the off-diagonal frequency blocks become a non-negligible part of the leading-order background covariance. Accordingly, $\chi_f\bigl(\bm R_c^{(0)}\bigr)$ indicates whether the complexity of the observation covariance is driven mainly by frequencywise energy variation or by medium-induced cross-frequency statistical coupling.

This interpretation follows directly from \eqref{eq:Rc0_block_entry}. The contrast field $\xi(\bm x,\omega_n)$ is obtained by mapping the same set of spatial random medium parameters through the Cole--Cole dispersive constitutive relation. Hence, if $C_{\xi}^{(2)}\bigl[(\bm x,\omega_n),(\bm x',\omega_{n'})\bigr]\neq 0$ for $n\neq n'$, the constitutive responses at different frequency points are not statistically independent, and the resulting cross-frequency coupling is transferred into the observation-domain covariance through the reference propagation kernels. When $\chi_f\bigl(\bm R_c^{(0)}\bigr)$ is negligible, the leading-order background covariance may be approximated as frequency block diagonal. Otherwise, the off-diagonal frequency covariance blocks are a substantive consequence of the random dispersive medium, and they cannot be omitted without loss of structural information.

\section{Experimental Validation and Result Analysis}\label{sec:experiments}

This section evaluates three aspects through numerical experiments: the first-order applicability boundary at the propagation level, the effectiveness of the cross-frequency structural metric $\chi_f$, and the structural consequences under unified scene settings.

Unless otherwise stated, the experiments in Sections~\ref{subsec:exp_propagation}--\ref{subsec:exp_structural_consequence} use the same discretization. The number of Monte Carlo samples is \num{256}, the number of frequency-diverse transmit channels is \num{6}, the number of receive channels is \num{6}, and the number of discrete patches is \num{96}. The frequency axis is fixed at \qtylist{50;70;90;110;130;150}{\mega\hertz}, corresponding to a low-frequency stepped-frequency GPR channelization used to expose medium-induced cross-frequency structure over a geophysically relevant band. The spatial grid is $8\times 12$, corresponding to $x\in[0,\qty{2.75}{\meter}]$ and $z\in[\qty{0.35}{\meter},\qty{2.10}{\meter}]$. The three representative medium scenes and their parameters are listed in Table~\ref{tab:cole_cole_parameters}.

For each scene, three reference states, denoted by $B$, $M$, and $U$, are considered. The state $B$ is the baseline reference that matches the nominal background and serves as the standard expansion point around the true scene. The state $M$ is a mismatched reference with an imposed deterministic offset and is used to assess the effect of reference mismatch on propagation, covariance, and structural metrics. The state $U$ is obtained from $B$ by reabsorbing the dominant deterministic bias according to the update strategy described in the paper, and is used to examine whether reference correction can restore the validity of the first-order expansion and the associated structural analysis.

\begin{table*}[!t]
	\centering
	\caption{Representative Cole--Cole parameter settings for the three benchmark scenarios}
	\label{tab:cole_cole_parameters}
	\begin{tblr}{
			width=\textwidth,
			colspec={Q[c,0.55] Q[l,1.85] Q[c,0.95] Q[c,0.95] Q[c,1.05] Q[c,1.0] Q[c,0.55] Q[c,0.8]},
			row{1}={font=\bfseries},
			cells={font=\footnotesize},
		}
		\toprule
		Code  & Scenario                                                                                                  & $\varepsilon_s$       & $\varepsilon_\infty$    & $\Delta\varepsilon$ & $\tau$ (\si{\second}) & $\alpha$    & $\sigma_{dc}$ (\si{\siemens\per\metre}) \\
		\midrule
		$S_1$ & Dry basalt / lava \cite{elshafie2013DielectricHardnessMeasurements,ahmad1990MagneticElectricalProperties}
		      & \numrange{8}{10}                                                                                          & \numrange{8}{10}      & $0$                     & \text{N/A}          & $0$                   & $\approx 0$                                           \\

		$S_2$ & Lunar regolith \cite{carrier1991PhysicalPropertiesLunar,olhoeft1974ElectricalPropertiesLunar}
		      & $1.919^{\rho}$                                                                                            & $1.919^{\rho}$        & $0$                     & \text{N/A}          & $0$                   & $\approx 0$                                           \\

		$S_3$ & Pure ice at \qty{0}{\celsius} \cite{auty1952DielectricPropertiesIce,evans1965DielectricPropertiesIce}
		      & \num{91.5}                                                                                                & \numrange{3.15}{3.17} & \numrange{88.33}{88.35} & \num{2.1e-5}        & $0$                   & $\approx 0$                                           \\
		\bottomrule
	\end{tblr}
\end{table*}

\subsection{Numerical Validation at the Propagation and Single-Channel Response Levels}\label{subsec:exp_propagation}

The propagation-level experiments address three questions. The first is whether the zeroth-order propagation skeleton is correctly established and responds clearly to changes in frequency, transmit channel, and receive channel. The second is the validity range of the distorted Born approximation under the present setting. The third is whether first-order closure holds from $\Delta\varepsilon$ to propagation perturbations and then to the single-channel background response.

\subsubsection{Identifiability of the Zeroth-Order Propagation Skeleton and Main Propagation Structure}

\begin{table*}[!t]
	\centering
	\caption{Energy scale and index sensitivity of the zero-order propagation skeleton}
	\label{tab:exp_prop_zero_order}
	\begin{tblr}{
		width=\textwidth,
		colspec={X[1]X[1]X[2]X[2]X[1.5]X[1.5]X[1.5]X[1.5]},
		row{1}={font=\bfseries\scriptsize},
		cells={font=\scriptsize},
		}

		\toprule
		Scenario & Reference & Mean $A_0$ energy & Peak $A_0$ energy & $\Delta_f(A_0)$ & $\Delta_r(A_0)$ & $\Delta_t(A_0)$ & $\overline{\Delta}(A_0)$ \\
		\midrule
		$S_1$    & B         & \num{3.0548e-7}   & \num{5.5773e-5}   & \num{1.6172}    & \num{3.1955}    & \num{3.1955}    & \num{2.7771}             \\
		$S_1$    & M         & \num{3.5060e-7}   & \num{6.1257e-5}   & \num{1.5916}    & \num{3.0999}    & \num{3.0999}    & \num{2.6464}             \\
		$S_1$    & U         & \num{3.0460e-7}   & \num{5.5625e-5}   & \num{1.6190}    & \num{3.2017}    & \num{3.1999}    & \num{2.7828}             \\
		$S_2$    & B         & \num{8.0263e-7}   & \num{1.6300e-4}   & \num{1.4478}    & \num{2.6850}    & \num{2.6850}    & \num{2.2878}             \\
		$S_2$    & M         & \num{7.4235e-7}   & \num{1.4942e-4}   & \num{1.4215}    & \num{2.7223}    & \num{2.7223}    & \num{2.2931}             \\
		$S_2$    & U         & \num{8.0111e-7}   & \num{1.6266e-4}   & \num{1.4472}    & \num{2.6846}    & \num{2.6843}    & \num{2.2870}             \\
		$S_3$    & B         & \num{7.8340e-7}   & \num{1.5833e-4}   & \num{1.4363}    & \num{2.6850}    & \num{2.6850}    & \num{2.2798}             \\
		$S_3$    & M         & \num{7.4117e-7}   & \num{1.4892e-4}   & \num{1.4206}    & \num{2.7161}    & \num{2.7161}    & \num{2.2881}             \\
		$S_3$    & U         & \num{9.0986e-8}   & \num{2.0052e-5}   & \num{1.4978}    & \num{3.1503}    & \num{3.1545}    & \num{2.6428}             \\
		\bottomrule
	\end{tblr}
\end{table*}

\begin{table}[!t]
	\centering
	\caption{Effect of reference-state variation on the zero-order propagation skeleton}
	\label{tab:exp_prop_reference_change}
	\begin{tblr}{
		width=\columnwidth,
		colspec={X[1]X[1]X[1.5]X[1.5]X[1.5]X[1.5]},
		row{1}={font=\bfseries\scriptsize},
		cells={font=\scriptsize},
		}

		\toprule
		Scenario & Reference & Rel. change in $G_t^{(0)}$ & Rel. change in $G_r^{(0)}$ & Rel. change in $A_0$ & Mean energy ratio of $A_0$ to $B$ \\
		\midrule
		$S_1$    & M         & \num{0.2050}               & \num{0.1753}               & \num{0.2292}         & \num{1.1477}                      \\
		$S_1$    & U         & \num{0.0240}               & \num{0.0191}               & \num{0.0209}         & \num{0.9971}                      \\
		$S_2$    & M         & \num{0.1371}               & \num{0.1159}               & \num{0.1668}         & \num{0.9249}                      \\
		$S_2$    & U         & \num{0.0120}               & \num{0.0098}               & \num{0.0138}         & \num{0.9981}                      \\
		$S_3$    & M         & \num{0.0948}               & \num{0.0800}               & \num{0.1154}         & \num{0.9461}                      \\
		$S_3$    & U         & \num{0.8759}               & \num{0.9121}               & \num{1.0566}         & \num{0.1161}                      \\
		\bottomrule
	\end{tblr}
\end{table}

Table~\ref{tab:exp_prop_zero_order} shows that the zeroth-order propagation skeleton $A_0$ responds clearly to scene, frequency, and channel variation. Stable stratification across scenes is already visible in the mean energy of $A_0$. Under the baseline reference state $B$, for example, the mean $A_0$ energy of $S_2$ and $S_3$ is \num{8.0263e-7} and \num{7.8340e-7}, respectively, which are about \num{2.63} and \num{2.56} times that of $S_1$. Different background media are therefore projected by the reference propagation kernels into clearly different zeroth-order structural scales, even before any first-order feedback is introduced.

The comparison among $\Delta_f(A_0)$, $\Delta_r(A_0)$, and $\Delta_t(A_0)$ further shows that the frequency-diverse channel sampling leaves an identifiable signature at the zeroth-order propagation level, although its effect is weaker than the structural change induced by switching a single transmit or receive channel. For $S_1$-B, $\Delta_r(A_0)=\Delta_t(A_0)=\num{3.1955}$, whereas $\Delta_f(A_0)=\num{1.6172}$. The same pattern appears in $S_2$ and $S_3$. At the same time, $\Delta_r(A_0)$ and $\Delta_t(A_0)$ remain nearly identical, indicating that the present discretization does not introduce an artificial transmit--receive imbalance at zeroth order.

Table~\ref{tab:exp_prop_reference_change} further shows that the effect of reference-state variation is scene dependent. For $S_1$ and $S_2$, the $U$ reference state induces only a small relative change in $A_0$ with respect to $B$, namely \numrange{0.0138}{0.0209}, while the mean-energy ratio remains close to \num{1}. In these two scenes, the update acts mainly as a local correction to the baseline reference. By contrast, for $S_3$ under the $U$ reference state, the relative change in $A_0$ reaches \num{1.0566} and the mean-energy ratio decreases to \num{0.1161}. The updated reference therefore no longer behaves as a small correction of the baseline skeleton.

\subsubsection{Validity Range of the Distorted Born Approximation}

\begin{table*}[!t]
	\centering
	\caption{Error and validity interval of the distorted Born approximation}
	\label{tab:exp_dba_validity}
	\begin{tblr}{
		width=\textwidth,
		colspec={X[0.5]X[0.5]X[1]X[1]X[1]X[1]X[1]X[1]X[1]},
		row{1}={font=\bfseries\scriptsize},
		cells={font=\scriptsize},
		}

		\toprule
		Scenario & Reference & $\bar e_{\mathrm{DBA}}(0.01)$ & $\bar e_{\mathrm{DBA}}(1.10)$ & $\mathcal A_{\mathrm{DBA}}$ & $s_{\mathrm{valid}}$ & $\mathcal M_{\mathrm{mono}}$ & $\mathcal A_{\Delta\varepsilon}$ & $\kappa_{\max}$ \\
		\midrule
		$S_1$    & B         & \num{8.0398e-10}              & \num{9.0816e-6}               & \num{1.1296e4}              & \num{1.10}           & \num{1.00}                   & \num{1.1000e2}                   & \num{1.0120}    \\
		$S_1$    & M         & \num{5.7993e-9}               & \num{6.9901e-5}               & \num{1.2053e4}              & \num{1.10}           & \num{1.00}                   & \num{1.1000e2}                   & \num{1.0234}    \\
		$S_1$    & U         & \num{6.0126e-10}              & \num{6.4998e-6}               & \num{1.0810e4}              & \num{1.10}           & \num{1.00}                   & \num{1.1000e2}                   & \num{1.0093}    \\
		$S_2$    & B         & \num{1.4770e-10}              & \num{9.0165e-7}               & \num{6.1046e3}              & \num{1.10}           & \num{1.00}                   & \num{1.1000e2}                   & \num{1.0032}    \\
		$S_2$    & M         & \num{1.2451e-9}               & \num{1.4518e-5}               & \num{1.1660e4}              & \num{1.10}           & \num{1.00}                   & \num{1.1000e2}                   & \num{1.0071}    \\
		$S_2$    & U         & \num{1.3844e-10}              & \num{7.2268e-7}               & \num{5.2200e3}              & \num{1.10}           & \num{1.00}                   & \num{1.1000e2}                   & \num{1.0030}    \\
		$S_3$    & B         & \num{1.0048e-9}               & \num{3.0274e-1}               & \num{3.0128e8}              & \num{0.70}           & \num{1.00}                   & \num{1.6068e4}                   & \num{4.8782}    \\
		$S_3$    & M         & \num{7.0663e-10}              & \num{2.9507e-1}               & \num{4.1757e8}              & \num{0.70}           & \num{1.00}                   & \num{1.8563e4}                   & \num{5.0660}    \\
		$S_3$    & U         & \num{6.1563e-2}               & \num{6.5739e-2}               & \num{1.0678}                & \num{1.10}           & \num{0.60}                   & \num{1.0348}                     & \num{5.2593}    \\
		\bottomrule
	\end{tblr}
\end{table*}

Table~\ref{tab:exp_dba_validity} summarizes the main boundaries of the distorted Born approximation. For all six combinations associated with $S_1$ and $S_2$, $\bar e_{\mathrm{DBA}}(0.01)$ remains at the \numrange{1e-10}{1e-9} level, and even at the strongest scan end $s=\num{1.10}$ it rises only to the \numrange{1e-7}{1e-5} level. All values of $s_{\mathrm{valid}}$ also reach the scan upper bound \num{1.10}. In these two weakly dispersive scenes, first-order distorted-Born truncation remains valid throughout the tested interval. Reference mismatch does increase the error constant, however. For example, $\bar e_{\mathrm{DBA}}$ for $S_1$-M reaches \num{6.9901e-5} at $s=\num{1.10}$, which is clearly larger than for $S_1$-B and $S_1$-U.

The approximation boundary appears much more clearly in $S_3$. For $S_3$-B and $S_3$-M, the error remains very small at the weak-perturbation end, but at $s=\num{1.10}$, $\bar e_{\mathrm{DBA}}$ rises abruptly to about \num{3e-1}, $\mathcal A_{\mathrm{DBA}}$ increases to \numrange{3e8}{4e8}, and $s_{\mathrm{valid}}$ is reduced to \num{0.70}. By contrast, $S_3$-U remains on a high plateau of about \num{6.5e-2} from the weakest-perturbation end onward, even though $\mathcal A_{\Delta\varepsilon}$ is only \num{1.0348}. This behavior is consistent with the substantial rewriting of the zeroth-order propagation skeleton in Table~\ref{tab:exp_prop_reference_change}. The $U$ reference state for $S_3$ has already moved the system away from the asymptotic regime centered on weak perturbations.

\subsubsection{First-Order Mapping from the Residual Dielectric Increment to Propagation Perturbations}

\begin{table*}[!t]
	\centering
	\caption{First-order mapping error from permittivity increment to propagation perturbation}
	\label{tab:exp_map_first_order}
	\begin{tblr}{
		width=\textwidth,
		colspec={X[0.8]X[0.7]X[1]X[1]X[1]X[1]X[1]X[1]X[1.3]X[1.2]},
		row{1}={font=\bfseries\scriptsize},
		cells={font=\scriptsize},
		}

		\toprule
		Scenario & Reference & $\bar e_G(0.01)$ & $\bar e_t(0.01)$ & $\bar e_G(1.10)$ & $\bar e_t(1.10)$ & $s_G^{\mathrm{valid}}$ & $s_t^{\mathrm{valid}}$ & $e_{\mathrm{side,max}}(1.10)$ & $\Delta_{tr}^{\mathrm{map}}$ \\
		\midrule
		$S_1$    & B         & \num{3.0090e-5}  & \num{3.0609e-5}  & \num{3.0e-3}     & \num{3.0e-3}     & \num{1.10}             & \num{1.10}             & \num{6.2e-3}                  & \num{1.0502e-13}             \\
		$S_1$    & M         & \num{8.0387e-5}  & \num{7.6879e-5}  & \num{8.8e-3}     & \num{8.4e-3}     & \num{1.10}             & \num{1.10}             & \num{1.8e-2}                  & \num{6.4259e-14}             \\
		$S_1$    & U         & \num{2.6095e-5}  & \num{2.6184e-5}  & \num{2.5e-3}     & \num{2.4e-3}     & \num{1.10}             & \num{1.10}             & \num{4.8e-3}                  & \num{2.7121e-13}             \\
		$S_2$    & B         & \num{2.7870e-5}  & \num{2.6686e-5}  & \num{9.1332e-4}  & \num{9.6677e-4}  & \num{1.10}             & \num{1.10}             & \num{1.9e-3}                  & \num{6.6154e-13}             \\
		$S_2$    & M         & \num{3.2694e-5}  & \num{3.4032e-5}  & \num{3.4e-3}     & \num{3.5e-3}     & \num{1.10}             & \num{1.10}             & \num{7.2e-3}                  & \num{2.4782e-14}             \\
		$S_2$    & U         & \num{2.6847e-5}  & \num{2.6939e-5}  & \num{9.0658e-4}  & \num{8.5546e-4}  & \num{1.10}             & \num{1.10}             & \num{1.7e-3}                  & \num{2.7611e-13}             \\
		$S_3$    & B         & \num{3.6247e-5}  & \num{3.5689e-5}  & \num{5.251e-1}   & \num{5.353e-1}   & \num{0.70}             & \num{0.70}             & \num{9.984e-1}                & \num{1.4064e-13}             \\
		$S_3$    & M         & \num{2.4094e-5}  & \num{2.3389e-5}  & \num{5.307e-1}   & \num{5.331e-1}   & \num{0.70}             & \num{0.70}             & \num{9.850e-1}                & \num{1.0817e-13}             \\
		$S_3$    & U         & \num{4.366e-1}   & \num{3.277e-1}   & \num{4.421e-1}   & \num{3.290e-1}   & --                     & --                     & \num{6.850e-1}                & \num{1.4803e-16}             \\
		\bottomrule
	\end{tblr}
\end{table*}

Table~\ref{tab:exp_map_first_order} directly evaluates the first-order mapping $\Delta\varepsilon\to\delta G,\delta G_t,\delta G_r$. For $S_1$ and $S_2$, the mean mapping errors at the weak-perturbation end are all at the \num{1e-5} level, and even at the strongest-perturbation end they increase only to \numrange{1e-3}{1e-2}. Correspondingly, both $s_G^{\mathrm{valid}}$ and $s_t^{\mathrm{valid}}$ reach \num{1.10}. The first-order propagation-level mapping is therefore numerically valid in these two weakly dispersive scenes.

Table~\ref{tab:exp_map_first_order} also shows that reference mismatch amplifies the mapping error. For example, in $S_1$, $\bar e_G$ for $S_1$-M reaches \num{8.8e-3} at the strongest-perturbation end, compared with \num{3.0e-3} and \num{2.5e-3} for $S_1$-B and $S_1$-U. This is consistent with the ordering of the DBA errors in Table~\ref{tab:exp_dba_validity}: reference-state error is transmitted continuously along the propagation chain.

Again, $S_3$ exhibits two distinct boundary mechanisms. For $S_3$-B and $S_3$-M, the mapping errors remain at the \num{1e-5} level at the weak-perturbation end, but rise abruptly to about \num{5.3e-1} at the strongest-perturbation end, and acceptable errors are maintained only for $s\le \num{0.70}$. By contrast, $S_3$-U starts from $\bar e_G=\num{0.4366}$ and $\bar e_t=\num{0.3277}$, and none of the three effective-intensity thresholds remains meaningful.

In addition, $\Delta_{tr}^{\mathrm{map}}$ stays between \num{1e-16} and \num{1e-13} for all nine combinations. Under the present discretization, the first-order mapping errors on the transmit and receive sides are therefore almost perfectly symmetric. The propagation-kernel mapping itself does not generate an imbalance in feedback-energy allocation between the two sides.

\subsubsection{Closure of the Single-Channel Background Response}

\begin{table*}[!t]
	\centering
	\caption{First-order closure of the single-channel background response}
	\label{tab:exp_single_channel_closure}
	\begin{tblr}{
		width=\textwidth,
		colspec={X[0.8]X[0.7]X[0.9]X[0.9]X[0.9]X[0.9]X[1]X[1]X[0.8]X[0.8]X[1.3]},
		row{1}={font=\bfseries\scriptsize},
		cells={font=\scriptsize},
		}

		\toprule
		Scenario & Reference & $\bar e_{\mathrm{main}}(0.01)$ & $\bar e_{\mathrm{1st}}(0.01)$ & $\bar e_{\mathrm{main}}(1.10)$ & $\bar e_{\mathrm{1st}}(1.10)$ & $\mathcal I_{\mathrm{1st}}(1.10)$ & $s_{\mathrm{1st}}^{\mathrm{valid}}$ & $\bar\eta_r$ & $\bar\eta_t$ & $\rho_{\times}(1.10)$ \\
		\midrule
		$S_1$    & B         & \num{6.9745e-5}                & \num{3.6764e-9}               & \num{7.6e-3}                   & \num{4.3000e-5}               & \num{1.7774e2}                    & \num{1.10}                          & \num{0.2520} & \num{0.7480} & \num{4.5426e-10}      \\
		$S_1$    & M         & \num{1.5682e-4}                & \num{2.2751e-8}               & \num{1.72e-2}                  & \num{2.7200e-4}               & \num{6.3221e1}                    & \num{1.10}                          & \num{0.3415} & \num{0.6585} & \num{2.0948e-8}       \\
		$S_1$    & U         & \num{5.8741e-5}                & \num{2.6897e-9}               & \num{6.5e-3}                   & \num{3.0982e-5}               & \num{2.0823e2}                    & \num{1.10}                          & \num{0.2375} & \num{0.7625} & \num{5.2389e-10}      \\
		$S_2$    & B         & \num{3.0328e-5}                & \num{4.6790e-10}              & \num{3.3e-3}                   & \num{4.2706e-6}               & \num{7.7907e2}                    & \num{1.10}                          & \num{0.2312} & \num{0.7688} & \num{4.4231e-12}      \\
		$S_2$    & M         & \num{7.1680e-5}                & \num{4.5846e-9}               & \num{7.9e-3}                   & \num{5.4470e-5}               & \num{1.4510e2}                    & \num{1.10}                          & \num{0.3427} & \num{0.6573} & \num{7.8501e-10}      \\
		$S_2$    & U         & \num{1.4617e-5}                & \num{3.6195e-10}              & \num{1.6e-3}                   & \num{2.5848e-6}               & \num{6.2236e2}                    & \num{1.10}                          & \num{0.2102} & \num{0.7898} & \num{1.8230e-12}      \\
		$S_3$    & B         & \num{7.4425e-5}                & \num{4.5362e-9}               & \num{8.650e-1}                 & \num{8.371e-1}                & \num{1.0333}                      & \num{0.70}                          & \num{0.1666} & \num{0.8334} & \num{0.3221}          \\
		$S_3$    & M         & \num{5.7181e-5}                & \num{2.2960e-9}               & \num{9.396e-1}                 & \num{1.0398}                  & \num{0.9036}                      & \num{0.70}                          & \num{0.3220} & \num{0.6780} & \num{0.9555}          \\
		$S_3$    & U         & \num{4.623e-1}                 & \num{3.250e-1}                & \num{4.949e-1}                 & \num{3.109e-1}                & \num{1.5917}                      & --                                  & \num{0.5000} & \num{0.5000} & \num{0.1363}          \\
		\bottomrule
	\end{tblr}
\end{table*}

Table~\ref{tab:exp_single_channel_closure} shows that first-order feedback is essential for closing the single-channel background response. In the two weakly dispersive scenes $S_1$ and $S_2$, first-order closure remains stable over the entire scan interval. When only the leading term is retained, $\bar e_{\mathrm{main}}(0.01)$ remains at the \num{1e-5} level. After first-order feedback is included, $\bar e_{\mathrm{1st}}(0.01)$ drops directly to the \numrange{1e-10}{1e-8} level. At the strongest-perturbation end, $\mathcal I_{\mathrm{1st}}(1.10)$ still remains within \numrange{6e1}{8e2}, and all values of $s_{\mathrm{1st}}^{\mathrm{valid}}$ reach \num{1.10}.

Table~\ref{tab:exp_single_channel_closure} also shows that reference mismatch changes the allocation of feedback terms between the two sides. Taking $S_2$ as an example, $\bar\eta_r$ for $S_2$-M increases relative to $S_2$-B and $S_2$-U. However, because $\rho_{\times}(1.10)$ remains very small, the second-order cross residual is still subordinate, and the closure quality degrades without collapsing.

The closure boundary appears again in $S_3$. For $S_3$-B and $S_3$-M, the closure-improvement factor remains at the \num{1e4} level at the weak-perturbation end, but at the strongest-perturbation end $\mathcal I_{\mathrm{1st}}(1.10)$ degrades to \num{1.0333} and \num{0.9036}, respectively, while $\rho_{\times}(1.10)$ increases to \num{0.3221} and \num{0.9555}. The second-order cross term is then no longer negligible, and first-order closure loses its dominance. By contrast, $S_3$-U never enters an acceptable closure interval over the entire scan range. The quantity $\bar e_{\mathrm{1st}}$ remains at the \num{3e-1} level, in full agreement with the previously observed rewriting of the propagation skeleton and the high-plateau mapping behavior.

Taken together, Tables~\ref{tab:exp_prop_zero_order}--\ref{tab:exp_single_channel_closure} show that, in $S_1$ and $S_2$, the zeroth-order propagation skeleton, distorted-Born truncation, first-order propagation mapping, and single-channel response closure form a mutually consistent chain. By contrast, $S_3$ consistently exhibits two types of boundary mechanism: abrupt failure at the strong-perturbation end and high-plateau failure after substantial rewriting of the reference skeleton.

\subsection{Validation, Attribution, and Applicability Boundaries of the Cross-Frequency Structural Metric}\label{subsec:exp_structure_metric}

On the basis of the covariance-level diagnosis, the cross-frequency coupling strength of the leading term is further examined,
\begin{equation}
	\chi_f\!\left(R_c^{(0)}\right)
	=
	\frac{\sum_{n\neq n^\prime}\norm{R^{(0)}_{c,nn^\prime}}_F^2}
	{\sum_n\norm{R^{(0)}_{c,nn}}_F^2}
	\label{eq:exp_chif}
\end{equation}
to determine whether it can indeed serve as a meaningful structural quantity. The validation around \eqref{eq:exp_chif} is carried out from three aspects: sensitivity to the definition itself, relation to the error of the block-diagonal approximation, and structural consequences under unified scene settings.

\subsubsection{Definition-Level Validation of the Cross-Frequency Coupling Strength}

\begin{table*}[!t]
	\centering
	\caption{Contrast of $\chi_f$ under zero-coupling and explicit-coupling constructions}
	\label{tab:exp_chif_definition}
	\begin{tblr}{
		colspec={X[2.2]X[1.2]X[1.7]X[1.7]X[1.6]X[1.6]},
		row{1}={font=\bfseries\footnotesize},
		cells={font=\footnotesize},
		}

		\toprule
		Construction                   & $\chi_f$     & Diagonal energy  & Off-diagonal energy & $\mathrm{err}_{\mathrm{bd,full}}$ & $\mathrm{err}_{\mathrm{bd,diag}}$ \\
		\midrule
		Zero-coupling construction     & \num{0.0156} & \num{5.0718e-10} & \num{7.9118e-12}    & \num{0.1239}                      & \num{0.1249}                      \\
		Explicit-coupling construction & \num{1.0534} & \num{5.8438e-10} & \num{6.1561e-10}    & \num{0.7162}                      & \num{1.0264}                      \\
		\bottomrule
	\end{tblr}
\end{table*}

Table~\ref{tab:exp_chif_definition} directly verifies the basic sensitivity of \eqref{eq:exp_chif}. The variation of $\chi_f$ is driven by substantive changes in the strength of the off-diagonal frequency blocks. Under the uncoupled construction, $\chi_f$ is only \num{0.0156}, and the corresponding off-diagonal energy is \num{7.9118e-12}. Under the explicitly coupled construction, $\chi_f$ increases to \num{1.0534}, and the off-diagonal energy becomes comparable to the diagonal energy.

Under the uncoupled construction, the residual nonzero value of $\chi_f$ also decreases monotonically as the number of Monte Carlo samples increases. When the sample size grows from \num{200} to \num{2000}, $\chi_f$ decreases from \num{0.0273} to \num{0.0025}. This shows that the quantity remains stable under sample convergence and can reliably indicate the presence of cross-frequency coupling.

\subsubsection{Relation to Medium Randomness}

\begin{table*}[!t]
	\centering
	\caption{Response of $\chi_f$ to global scaling and correlation length}
	\label{tab:exp_chif_mechanism}
	\begin{tblr}{
		width=\textwidth,
		colspec={X[1.2]X[1.4]X[1.4]X[1.4]X[1.5]X[1.5]X[1.6]},
		row{1}={font=\bfseries\scriptsize},
		cells={font=\scriptsize},
		}

		\toprule
		Scenario & $\chi_f$ at min scale & $\chi_f$ at max scale & Trace amplification & $\chi_f$ at min corr. length & $\chi_f$ at max corr. length & Corr. length ratio \\
		\midrule
		$S_1$    & \num{1.7612}          & \num{1.7612}          & \num{12100}         & \num{1.1756}                 & \num{2.5442}                 & \num{2.1642}       \\
		$S_2$    & \num{2.5966}          & \num{2.5966}          & \num{12100}         & \num{1.6890}                 & \num{3.6986}                 & \num{2.1898}       \\
		$S_3$    & \num{2.7524}          & \num{2.7524}          & \num{12100}         & \num{2.0352}                 & \num{3.2101}                 & \num{1.5773}       \\
		\bottomrule
	\end{tblr}
\end{table*}

\begin{table}[!t]
	\centering
	\caption{Sensitivity of $\chi_f$ to parameter channels}
	\label{tab:exp_chif_param}
	\begin{tblr}{
		width=\columnwidth,
		colspec={X[4]X[2]X[2]X[2]},
		row{1}={font=\bfseries\footnotesize},
		cells={font=\footnotesize},
		}

		\toprule
		Scenario--parameter           & Min. $\chi_f$ & Max. $\chi_f$ & Range            \\
		\midrule
		$S_1$ -- $\varepsilon_\infty$ & \num{1.6980}  & \num{1.6980}  & \num{9.1038e-15} \\
		$S_2$ -- $\varepsilon_\infty$ & \num{2.5496}  & \num{2.5496}  & \num{1.5543e-14} \\
		$S_3$ -- $\varepsilon_\infty$ & \num{2.8895}  & \num{2.8895}  & \num{2.1316e-14} \\
		$S_3$ -- $\Delta\varepsilon$  & \num{1.9310}  & \num{1.9310}  & \num{6.7042e-12} \\
		$S_3$ -- $\tau$               & \num{1.7399}  & \num{1.7452}  & \num{5.3e-3}     \\
		\bottomrule
	\end{tblr}
\end{table}

Table~\ref{tab:exp_chif_mechanism} separates structural variation from pure energy scaling. For all three scenes, $\chi_f$ remains essentially unchanged while the overall intensity is scanned from \num{0.01} to \num{1.10}, whereas $\operatorname{tr}(R_c^{(0)})$ is amplified by a factor of \num{12100}. Pure amplitude scaling therefore changes the power level but not the structural proportion measured by $\chi_f$, namely the relative allocation between off-diagonal and diagonal frequency blocks.

The correlation-length scan in the same table gives a different response. For $S_1$, $S_2$, and $S_3$, $\chi_f$ all increase significantly with correlation length, and the ratio reaches \num{2.1898} for $S_2$. Longer-range spatial correlation therefore appears more likely to preserve common random mechanisms across different frequency channels through the dispersive mapping, thereby strengthening cross-frequency coupling in the observation domain.

Table~\ref{tab:exp_chif_param} further narrows the physical interpretation. In the present implementation, separate scans of $\varepsilon_\infty$ and $\Delta\varepsilon$ produce almost no change in $\chi_f$, whereas the clearly identifiable influence arises from the relaxation parameter $\tau$ in $S_3$. This is consistent with the physical behavior of strongly relaxing scenes: varying $\tau$ changes how different frequencies share the same dispersive transition interval, and thus more directly affects cross-frequency statistical coupling.

\subsubsection{Relation to the Error of Block-Diagonal Approximation}

\begin{table}[!t]
	\centering
	\caption{Exact correspondence between $\chi_f$ and block-diagonal approximation error}
	\label{tab:exp_chif_blockdiag}
	\begin{tblr}{
		width=\columnwidth,
		colspec={X[7.2]X[2.8]},
		row{1}={font=\bfseries\footnotesize},
		cells={font=\footnotesize},
		}

		\toprule
		Statistic                                                             & Value            \\
		\midrule
		corr$(\chi_f,\mathrm{err}_{\mathrm{bd,full}})$                        & \num{0.97823}    \\
		corr$(\chi_f,\mathrm{err}_{\mathrm{bd,diag}})$                        & \num{0.99765}    \\
		corr$(\chi_f,\mathrm{err}_{\mathrm{bd,full}}^2)$                      & \num{0.98246}    \\
		corr$(\chi_f,\mathrm{err}_{\mathrm{bd,diag}}^2)$                      & \num{1.00000}    \\
		max residual of $\mathrm{err}_{\mathrm{bd,diag}}^2-\chi_f$            & \num{1.3323e-15} \\
		max residual of $\mathrm{err}_{\mathrm{bd,full}}^2-\chi_f/(1+\chi_f)$ & \num{6.6613e-16} \\
		\bottomrule
	\end{tblr}
\end{table}

Table~\ref{tab:exp_chif_blockdiag} shows an almost exact numerical correspondence between $\chi_f$ and the error of the block-diagonal approximation. In particular, corr$(\chi_f,\mathrm{err}_{\mathrm{bd,diag}}^2)$ reaches \num{1}, while the maximum residual of $\mathrm{err}_{\mathrm{bd,diag}}^2-\chi_f$ is only \num{1.3323e-15}, which is already at machine precision. By definition, $\chi_f$ directly determines how much off-diagonal frequency structure is lost under the block-diagonal approximation in the Frobenius sense.

\subsubsection{Boundary as a Proxy for the Full Response Structure}

\begin{table*}[!t]
	\centering
	\caption{Validity range of the main structural metric as a proxy for the full response}
	\label{tab:exp_chif_proxy}
	\begin{tblr}{
		width=\textwidth,
		colspec={X[1.2]X[1]X[1.4]X[1.4]X[1.6]X[1.8]X[1.6]},
		row{1}={font=\bfseries\scriptsize},
		cells={font=\scriptsize},
		}

		\toprule
		Scenario & Scale      & $\chi_f$ (main) & $\chi_f$ (full) & $\delta_{\chi,\mathrm{rel}}$ & Proxy error (Frobenius) & Valid \\
		\midrule
		$S_1$    & \num{0.01} & \num{1.7860}    & \num{1.7860}    & \num{4.0457e-6}              & \num{8.0131e-6}         & Yes   \\
		$S_1$    & \num{0.30} & \num{1.7860}    & \num{1.7858}    & \num{1.2147e-4}              & \num{2.4013e-4}         & Yes   \\
		$S_1$    & \num{1.10} & \num{1.7860}    & \num{1.7852}    & \num{4.4632e-4}              & \num{8.7781e-4}         & Yes   \\
		$S_2$    & \num{0.01} & \num{2.6434}    & \num{2.6434}    & \num{1.8989e-6}              & \num{6.5443e-6}         & Yes   \\
		$S_2$    & \num{0.30} & \num{2.6434}    & \num{2.6435}    & \num{5.6868e-5}              & \num{1.9625e-4}         & Yes   \\
		$S_2$    & \num{1.10} & \num{2.6434}    & \num{2.6439}    & \num{2.0752e-4}              & \num{7.1875e-4}         & Yes   \\
		$S_3$    & \num{0.01} & \num{2.8552}    & \num{2.8551}    & \num{1.0407e-5}              & \num{1.3257e-4}         & Yes   \\
		$S_3$    & \num{0.30} & \num{2.8551}    & \num{2.8551}    & \num{1.2756e-6}              & \num{1.3487e-4}         & Yes   \\
		$S_3$    & \num{1.10} & \num{1.3697}    & \num{1.2791}    & \num{7.08e-2}                & \num{7.598e-1}          & No    \\
		\bottomrule
	\end{tblr}
\end{table*}

Table~\ref{tab:exp_chif_proxy} gives the applicability boundary for using the leading-order structural quantity as a proxy for the full response structure. In $S_1$ and $S_2$, both $\delta_{\chi,\mathrm{rel}}$ and the Frobenius proxy error remain very small at all three perturbation scales. The leading-order $\chi_f$ is therefore sufficient to represent the dominant cross-frequency structure of the full response. Although the leading-order covariance does not fully close the entire response with high accuracy, it still provides a stable proxy for the dominant structural quantity.

For $S_3$, the leading-order proxy remains valid at scales \num{0.01} and \num{0.30}. When the scale increases to \num{1.10}, however, $\delta_{\chi,\mathrm{rel}}$ rises to \num{7.08e-2} and the Frobenius proxy error reaches \num{7.598e-1}. Under strong relaxation and large perturbation, the full response structure has already been reshaped by the feedback terms, and the leading-order $\chi_f$ alone is no longer sufficient to describe the overall cross-frequency structure.

Taken together, these results show that $\chi_f$ is sensitive to off-diagonal frequency blocks, stable under pure energy scaling, clearly responsive to correlation length and the relaxation parameter, and directly linked to the error of the block-diagonal approximation. Its role as a proxy for the full response, however, remains regime dependent, and in particular breaks down at the strong-perturbation end of $S_3$.

\subsection{Reference State, Scene Differences, and Structural Consequences}\label{subsec:exp_structural_consequence}

Based on the above results, this subsection addresses three questions: whether reselection of the reference state improves the stability of the structural metric, what structural differences appear across different physical scenes under unified geometry and randomness, and whether cross-frequency statistical structure produces observable minimal processing consequences.

\subsubsection{Effect of Reference-State Reselection on Structural Stability}

\begin{table*}[!t]
	\centering
	\caption{Influence of reference-state selection on response error and structural stability}
	\label{tab:exp_reference_update}
	\begin{tblr}{
		width=\textwidth,
		colspec={X[0.8]X[0.7]X[1]X[1.1]X[1]X[1]X[0.9]X[1.2]X[1.1]X[1.2]},
		row{1}={font=\bfseries\scriptsize},
		cells={font=\scriptsize},
		}

		\toprule
		Scenario & Reference & Mean channel error & Covariance error & $\chi_{\mathrm{main}}$ & $\chi_{\mathrm{rich}}$ & $\abs{\Delta\chi}$ & Mean $\norm{\Delta\varepsilon}$ & Mean $\eta_{\mathrm{ref}}$ & $P(\eta_{\mathrm{ref}}<1)$ \\
		\midrule
		$S_1$    & B         & \num{0.0040}       & \num{7.5504e-4}  & \num{1.7523}           & \num{1.7514}           & \num{8.7278e-4}    & \num{8.5611e-11}                & \num{0.0431}               & \num{1}                    \\
		$S_1$    & M         & \num{0.0111}       & \num{1.30e-2}    & \num{1.9090}           & \num{1.9035}           & \num{5.5e-3}       & \num{2.4499e-10}                & \num{0.1259}               & \num{1}                    \\
		$S_1$    & U         & \num{0.0036}       & \num{1.5310e-4}  & \num{1.7428}           & \num{1.7427}           & \num{3.1911e-5}    & \num{7.6489e-11}                & \num{0.0381}               & \num{1}                    \\
		$S_2$    & B         & \num{0.0020}       & \num{6.9056e-4}  & \num{2.5140}           & \num{2.5144}           & \num{3.7967e-4}    & \num{2.1940e-11}                & \num{0.0111}               & \num{1}                    \\
		$S_2$    & M         & \num{0.0059}       & \num{8.3e-3}     & \num{2.5317}           & \num{2.5283}           & \num{3.4e-3}       & \num{8.0293e-11}                & \num{0.0407}               & \num{1}                    \\
		$S_2$    & U         & \num{0.0017}       & \num{1.3499e-4}  & \num{2.5003}           & \num{2.5003}           & \num{1.1861e-5}    & \num{1.9307e-11}                & \num{0.0096}               & \num{1}                    \\
		$S_3$    & B         & \num{0.5542}       & \num{0.7292}     & \num{2.9782}           & \num{2.6343}           & \num{0.3440}       & \num{1.0895e-8}                 & \num{5.4643}               & \num{0}                    \\
		$S_3$    & M         & \num{0.5493}       & \num{0.7132}     & \num{2.9861}           & \num{2.7246}           & \num{0.2616}       & \num{1.0864e-8}                 & \num{5.4458}               & \num{0}                    \\
		$S_3$    & U         & \num{0.5235}       & \num{0.6713}     & \num{1.5996}           & \num{1.0800}           & \num{0.5196}       & \num{1.0814e-8}                 & \num{5.4203}               & \num{0}                    \\
		\bottomrule
	\end{tblr}
\end{table*}

Table~\ref{tab:exp_reference_update} shows that updating the reference state can improve both response accuracy and structural stability in weakly dispersive scenes. For $S_1$, the covariance error decreases from \num{1.30e-2} under the $M$ reference state to \num{1.5310e-4} under the $U$ reference state. For $S_2$, the corresponding quantity decreases from \num{8.3e-3} to \num{1.3499e-4}. At the same time, $\abs{\Delta\chi}$ also decreases sharply. The $U$ reference state therefore reduces the response error and restabilizes the leading-order structural metric.

The proxy quantity $\eta_{\mathrm{ref}}$ is highly consistent with this improvement. For $S_1$ and $S_2$, all three reference states satisfy $P(\eta_{\mathrm{ref}}<1)=1$, indicating that the residual dielectric increment remains within a regime in which first-order feedback can dominate. The $U$ reference state moves the system further toward the center of this stable regime. In contrast, for $S_3$, all three reference states satisfy $P(\eta_{\mathrm{ref}}<1)=0$, and the mean $\eta_{\mathrm{ref}}$ remains about \num{5.4}. Even though the $U$ reference state is slightly better than $B$ on some local indicators, it does not alter the underlying deviation mechanism in the strongly relaxing scene.

\subsubsection{Structural Differences Across Unified Scenes and Their Minimal Processing Consequences}

\begin{table*}[!t]
	\centering
	\caption{Main-term structural differences across the three scenarios under a unified setting}
	\label{tab:exp_scene_compare}
	\begin{tblr}{
		width=\textwidth,
		colspec={X[1.2]X[1.5]X[1.8]X[1.5]X[1.3]X[1.4]X[1.3]},
		row{1}={font=\bfseries\scriptsize},
		cells={font=\scriptsize},
		}

		\toprule
		Scenario & $\chi_f^{(\xi)}$ & $\chi_f(R_c^{(0)})$ & Block-diag. error & $\kappa_{\mathrm{prop}}$ & $\operatorname{tr}(R_c^{(0)})$ & $\norm{R_c^{(0)}}_F$ \\
		\midrule
		$S_1$    & \num{5.0000}     & \num{1.6875}        & \num{0.7924}      & \num{0.3375}             & \num{2.2203e-8}                & \num{1.7942e-8}      \\
		$S_2$    & \num{5.0000}     & \num{2.6404}        & \num{0.8516}      & \num{0.5281}             & \num{5.1382e-8}                & \num{4.0073e-8}      \\
		$S_3$    & \num{5.0000}     & \num{3.1591}        & \num{0.8715}      & \num{0.6318}             & \num{2.8620e-3}                & \num{2.2000e-3}      \\
		\bottomrule
	\end{tblr}
\end{table*}

Table~\ref{tab:exp_scene_compare} shows that the three scenes differ mainly in how the propagation layer preserves constitutive cross-frequency coupling. The values of $\chi_f^{(\xi)}$ are all close to \num{5} and are therefore barely distinguishable at the constitutive level. Once projected into the leading-order observation covariance, however, $\chi_f(R_c^{(0)})$ separates clearly in the order $S_1<S_2<S_3$, and the corresponding $\kappa_{\mathrm{prop}}$ increases from \num{0.3375} to \num{0.6318}. The scene differences are therefore shaped primarily by the propagation layer through how strongly it preserves or compresses the coupling.

The block-diagonal error in Table~\ref{tab:exp_scene_compare} follows the same ordering $S_1<S_2<S_3$, which is fully consistent with the interpretation of $\chi_f$. Although $\operatorname{tr}(R_c^{(0)})$ for $S_3$ is already much larger than in the other two scenes, the degree of failure of the block-diagonal approximation is still governed by the structural differences described by $\chi_f$ and $\kappa_{\mathrm{prop}}$.

\begin{table}[!t]
	\centering
	\caption{Influence of covariance-structure simplification on minimal processing outcomes}
	\label{tab:exp_minimal_consequence}
	\begin{tblr}{
		width=\columnwidth,
		colspec={X[1.2]X[2]X[1.8]X[1.8]X[1.8]X[1.4]},
		row{1}={font=\bfseries\footnotesize},
		cells={font=\footnotesize},
		}

		\toprule
		Scenario & Covariance structure & $\chi_f^{\mathrm{ext}}$ & Whitening error & Subspace capture & $p_{0.9}$ \\
		\midrule
		$S_1$    & Full                 & \num{1.7845}            & \num{0.7284}    & \num{0.9646}     & \num{2}   \\
		$S_1$    & Block                & \num{0}                 & \num{1.8124}    & \num{0.6256}     & \num{2}   \\
		$S_1$    & Diag                 & \num{0}                 & \num{3.9450}    & \num{0.3491}     & \num{2}   \\
		$S_2$    & Full                 & \num{2.4818}            & \num{0.7600}    & \num{0.9766}     & \num{2}   \\
		$S_2$    & Block                & \num{0}                 & \num{1.9276}    & \num{0.4466}     & \num{2}   \\
		$S_2$    & Diag                 & \num{0}                 & \num{4.0217}    & \num{0.2817}     & \num{2}   \\
		$S_3$    & Full                 & \num{2.8460}            & \num{0.2373}    & \num{0.9388}     & \num{2}   \\
		$S_3$    & Block                & \num{0}                 & \num{1.8616}    & \num{0.4803}     & \num{2}   \\
		$S_3$    & Diag                 & \num{0}                 & \num{3.9298}    & \num{0.2289}     & \num{2}   \\
		\bottomrule
	\end{tblr}
\end{table}

If the cross-frequency covariance blocks are structurally meaningful, then progressively simplifying the full-frequency covariance into a block-diagonal or even channelwise independent form must produce observable consequences in whitening and principal-subspace extraction. Table~\ref{tab:exp_minimal_consequence} confirms this in all three scenes. Taking $S_2$ as an example, the whitening error increases from \num{0.7600} for Full to \num{1.9276} for Block and then to \num{4.0217} for Diag. The subspace capture decreases from \num{0.9766} to \num{0.4466} and then to \num{0.2817}. Removing only the cross-frequency covariance blocks is already sufficient to degrade the most basic processing results substantially, and removing within-frequency channel correlation causes a further decline.

The scene ordering in Table~\ref{tab:exp_minimal_consequence} is also informative. $S_3$ has the largest $\chi_f^{\mathrm{ext}}$, and the increase in whitening error from Full to Block is also the most pronounced. Cross-frequency statistical information in strongly relaxing scenes is therefore more difficult to replace with simplified covariance models.

\section{Conclusion}\label{sec:conclusion}

This paper studies background-clutter structure induced by random dispersive media in single-snapshot FDA-MIMO-GPR. The analysis connects constitutive mapping, propagation response, and cross-frequency structural quantities within a unified framework. Representative media are modeled by the Cole--Cole formulation. Under a reference-medium framework, a normalized incremental contrast function is constructed, a single-snapshot background-response expression retaining first-order propagation-kernel feedback is derived, and a cross-frequency coupling strength of the leading-order background covariance is defined. This quantity captures the relative strength of off-diagonal frequency covariance blocks and clarifies how random medium fluctuations are written into observation-domain cross-frequency structure.

Numerical results show strong consistency in weakly dispersive scenes across constitutive mapping, the zeroth-order propagation skeleton, distorted-Born truncation, first-order propagation mapping, and single-channel response closure. The proposed structural metric distinguishes uncoupled and explicitly coupled constructions, is stable under pure energy scaling, responds clearly to correlation length and relaxation, and corresponds directly to the error of the frequency block-diagonal approximation. Further experiments show that this cross-frequency structure has observable consequences for whitening and principal-subspace extraction.

The present theory has a clear scope. Under the current semi-nonlinear model, the background-response covariance is not generally determined by the second-order statistics of the contrast field alone. Strongly relaxing scenes exhibit two recurring boundary behaviors: abrupt failure at the strong-perturbation end and a high-error plateau after substantial rewriting of the reference skeleton. The framework is therefore mainly valid when the residual dielectric increment relative to the reference medium remains within the first-order-feedback regime. Outside this regime, propagation feedback and higher-order statistics become important, and the leading-order structural quantity no longer provides a sufficient proxy for the full response structure.

Future work may extend the present framework through adaptive reference updating, higher-order propagation corrections, and richer statistical closure, and may further assess the engineering value of the proposed structural quantity in downstream single-snapshot processing.

\bibliographystyle{IEEEtran}
\bibliography{ref}
\end{document}